\documentclass{article}                     
\usepackage[margin=2cm]{geometry}
\usepackage{graphicx}

\usepackage[utf8]{inputenc}
\usepackage{amsfonts,amssymb,amsmath}
\usepackage{color}

\newtheorem{theorem}{Theorem}
\newtheorem{proposition}{Proposition}

\newtheorem{corollary}{Corollary}
\newtheorem{remark}{Remark}
\graphicspath{{figures/}}

\usepackage{color}
\def\inv{{^{-1}}}

\def\RR{{\mathbb R}}

\def\CC{\mathbb C}
\def\ZZ{\mathbb Z}

\def\Id{{\mathbbm 1}}

\def\tX{\tilde{X}}

\def\be{\begin{equation}}
\def\beq#1{\begin{equation}\label{#1}}
\def\ee{\end{equation}}
\def\bea{\begin{eqnarray}}
\def\beqa#1{\begin{eqnarray}\label{#1}}
\def\eea{\end{eqnarray}}
\def\ba{\begin{array}}
\def\ea{\end{array}}

\DeclareMathAlphabet{\mathpzc}{OT1}{pzc}{m}{it}

\def\cA{{\mathcal A}}

\def\cD{{\mathcal D}}

\def\cH{{\mathcal H}}

\def\cU{{\mathcal U}}
\def\cV{{\mathcal V}}

\def\R{\RR}

\def\Ltwo{L^2(\RR )}

\def\Htwo{H^2(\RR )}

\def\bds{_{-\infty}^\infty}

\def\ba{{\mathbf a}}

\def\tf{{\tilde{f}}}

\def\tX{{\tilde{X}}}

\newcommand{\tP}{\widetilde{P}}
\renewcommand{\Id}{{\bf 1}}
\newcommand{\tcU}{\widetilde{\cU}}

\begin{document}

\title{A survey of uncertainty principles and some signal processing applications\thanks{This work was supported by the European project UNLocX, grant n. 255931. B. Ricaud is with the      Signal Processing Laboratory 2,  Ecole Polytechnique F\'ed\'erale de Lausanne (EPFL), Station 11, 1015 Lausanne, Switzerland. B. Torr\'esani is with the LATP, Aix-Marseille Univ/CNRS/Centrale Marseille, UMR7353, 39 rue Joliot-Curie, 13453 Marseille cedex 13, France.}
}


\author{Benjamin Ricaud         \and       Bruno Torr\'esani
}




\maketitle

\begin{abstract}
The goal of this paper is to review the main trends in the domain of uncertainty
principles and localization, highlight their mutual connections and
investigate practical consequences. The discussion
is strongly oriented towards, and motivated by signal processing problems, from
which significant advances have been made recently.
Relations with sparse
approximation and coding problems are emphasized.

\end{abstract}


\section{Introduction}
\label{intro}
Uncertainty inequalities generally express the impossibility for a function (or
a vector in the discrete case) to be simultaneously {\em sharply concentrated}
in two different representations, provided the latter are {\em incoherent
  enough}. Such a loose definition can be made concrete by further specifying
the following main ingredients:
\begin{itemize}
\item
\textbf{A global setting}, generally a couple of Hilbert spaces (of functions or
vectors) providing two representations for the objects of interest (e.g. time
and frequency, or more general phase space variables).
\item \textbf{An invertible linear transform} (operator, matrix) mapping the
  initial representation to the other one, without information loss.
\item
\textbf{A concentration measure} for the elements of the two representation spaces:
variance, entropy, $L^p$ norms,...
\end{itemize}
Many such settings have been proposed in the literature during
the last century, for various purposes. The first formulation was proposed in
quantum mechanics where the uncertainty principle is still a major concern.
However it is not restricted to this field and appears whenever one has to
represent functions and vectors in different manners, to extract some specific
information. This is basically what is done in signal processing where the
uncertainty principle is of growing interest. 

The basic (quantum mechanical) prototype is provided by the so-called
Robertson-Schr\"odinger inequality, which establishes a lower bound for the
product of variances of any two self-adjoint operators on a generic Hilbert
space.  The most common version of the principle is as follows:
\begin{theorem}
\label{th:robertson}
Let $f\in\cH$ (Hilbert space), with $\|f\|=1$. Let $A$ and $B$ be (possibly
unbounded) self-adjoint operators on $\cH$ with respective domains $D(A)$ and
$D(B)$. Define the mean and variance of $A$ in state 
$f\in D(A)$ by
$$
e_f(A) = \langle Af,f\rangle\ ,\qquad v_f(A) = e_f(A^2) - e_f(A)^2\ .
$$
Setting  $[A,B]=AB-BA$ and $\{A,B\}=AB+BA$, we have
$\forall f\in D(AB)\cap D(BA)$,
$$
v_f(A)v_f(B)\ge \frac1{4}\left[
|e_f([A,B])|^2 + |e_f( \{A- e_f(A),B- e_f(B)\})|^2\right]\ .
$$
\end{theorem}
The quantities $v_f(A)$ and $v_f(B)$ can also be interpreted as the variances of
two representations of $f$ given by its projection onto the
respectives bases of (possibly generalized) eigenvectors of $A$ and $B$.
From the self-adjointness of $A$ and $B$, there exists
a unitary operator mapping one representation to the other.

The proof of this result is quite generic and carries over many
situations. However, the choice of the variance to measure concentration
properties may be quite questionable in a number of practical situations, and
several alternatives have been proposed and studied in the literature.

The goal of this paper is to summarize a part of the literature on this
topic, discuss a few recent results and focus on specific signal processing
applications.
We shall first describe the continuous setting, before moving to discrete
formulations and emphasizing the main differences. Given the space limitations,
the current paper cannot be exhaustive. We have selected a few examples which highlight the structure and some important aspects of the uncertainty principle. We refer for example
to~\cite{Folland97uncertainty} for a very good and complete account of classical
uncertainty relations, focused on time-frequency uncertainty. An information theory point of view of the uncertainty principle may be found in~\cite{Dembo91information} 
and a review of entropic uncertainty principles has
been given in~\cite{Wehner10entropic}. More recent contributions, mainly in the
sparse approximation literature, introducing new localization measures will be
mentioned in the core of the paper.


\section{Some fundamental aspects of the uncertainty principle}
\subsection{Signal representations}\label{signalrep}
The uncertainty principle is usually understood as a relation between the
simultaneous spreadings of a function and its Fourier transform. More generally,
as expressed in Theorem~\ref{th:robertson} an uncertainty principle also
provides a relation between any two representations of a function, here the ones
given by its projections onto the (possibly generalized) eigenbases of $A$ and $B$.
Can a representation be something else than the projection onto a (generalized)
eigenbasis? The answer is yes: representations can be made by
introducing frames. A set of vectors  $\cU=\{u_k\}_k$ in a  Hilbert space $\cH$
is a frame of $\cH$ if for all $f\in\cH$:
\begin{align}
A\|f\|^2\le\sum_k|\langle f,u_k \rangle|^2\le B\|f\|^2,
\end{align}
where $A$, $B$ are two constants such that $0<A\le B<\infty$. Since $A>0$, any
$f\in\cH$ can be recovered from its frame coefficients $\{\langle f,u_k\rangle\}_k$.
This is a key point: in order to
compare two representations, information must not be lost in the process.
Orthonormal bases are particular cases of frames for which $A=B=1$ and the frame
vectors are orthogonal.

Denote by $U: f\in\cH\to \{\langle f,u_k\rangle\}_k$ the so-called analysis
operator. $U$ is left invertible, which yields inversion formulas of the form
$$
f = \sum_k \langle f,u_k\rangle \tilde u_k\ ,
$$
where $\tcU=\{\tilde u_k\}_k$ is an other family of vectors in $\cH$, which can also
be shown to be a frame, termed dual frame of $\cU$. Choosing as left inverse the
Moore-Penrose pseudo-inverse $U\inv=U^\dag$ yields the so called canonical dual
frame $\tcU^\circ=\{\tilde u_k^\circ=(UU^*)\inv u_k\}_k$, but other choices are
possible.

The uncertainty principle can be naturally extended to frame representations,
i.e. representations of vectors $f\in\cH$ by their frame coefficients.
As before, uncertainty inequalities limit the extend to which a vector can have
two arbitrarily concentrated frame representations. Since variances are not
necessarily well defined in such a case, other concentrations measures such as
entropies have to be used. For example, bounds for the entropic uncertainty
principle are derived in~\cite{RicaudTorresani}.

\subsection{The mutual coherence: how different are two representations ?}
A second main aspect of uncertainty inequalities is the heuristic remark that
{\em the more different the representations, the more constraining the bounds}.
However, one needs to be able to measure how different two representations are.
This is where the notion of coherence enters. 

Let us first stick to frames in the discrete setting. Let $\cU=\{u_k\}_k$ and
$\cV=\{v_k\}_k$ be two frames of $\cH$.
Let us define the operator $T=VU^{-1}$ which allows one to pass from the
representation of $f$ in $\cU$ to the one in $\cV$. It is given by:
$$
TUf(j)= \left\langle\sum_k(Uf)(k)\,\tilde{u}_k,v_j\right\rangle=
\sum_k (Uf)(k)\,\langle \tilde{u}_k,v_j\rangle . 
$$
This relation shows that in finite dimension $T$ is represented by a matrix
$G=G(\tcU,\cV)$ (the cross Gram matrix of $\tcU$ and $\cV$) defined by
$G_{j,k}=\langle \tilde{u}_k,v_j\rangle$. The matrix $G$ encodes the differences
of the two frames. The latter can be measured by various norms of $T$, among which
the so-called {\em mutual coherence}:
\beq{fo:mutual.coherence}
\mu=\mu(\tcU,\cV)=\max_{j,k}| \langle\tilde{u}_k,v_j\rangle|=\max_{j,k}| G_{j,k}| =
\|T\|_{\ell^1\to\ell^\infty}\ .
\ee
This quantity encodes (to some extend) the algebraic properties of $T$.
\begin{remark}
This particular quantity (norm) may be generalized to other kinds of norms which
would be, depending on the setting, more appropriate for the estimation of the
correlation between the two representations. Indeed, it is the characterization
of the matrix which quantify how close two representations actually are.
 \end{remark}
 \begin{remark}
 In the standard case ($N$-dimensional) where the uncertainty is stated between
 the Kronecker and Fourier bases, $|T_{j,k}|=1/\sqrt{N}$ for all $j,k$. These
 bases are said to be mutually unbiased and $\mu=1/\sqrt{N}$ is the smallest
 possible value of $\mu$.
 \end{remark}

In the case of the entropic uncertainty principle, the demonstration of the
inequality is based on the Riesz interpolation theorem and it rely on bounds of
$T$ as an operator from $\ell^1\to\ell^\infty$ and from $\ell^2\to\ell^2$(see
section~\ref{se:sparse}). 
As we shall see, this notion of mutual coherence appears in most of the
uncertainty relations. A noticeable exception is the variance-based uncertainty
principle. In this case it is replaced by the commutation relation between the
two self-adjoint operators and the connection with the coherence is not
straightforward.

\subsection{The notion of phase space}
Standard uncertainty principles are associated with pairs of representations:
time localization vs frequency localization, time localization vs scale
localization,... However, in some situations, it is possible to introduce
directly a {\em phase} space, which involves jointly the two representation
domains, in which (non-separable) uncertainty principles can be directly
formulated: joint time-frequency space, joint time-scale space,...

Uncertainty principles associated with pairs of representations often have
counterparts defined directly in the joint space. We shall see a few examples in
the course of the current paper. In such situations, the mutual coherence is
replaced with a notion of phase space coherence.

\section{Uncertainty inequalities in continuous settings: a few remarkable examples}
\label{se:continuous}
To get better insights on the uncertainty principle we state here a few
remarkable results which illustrate the effect of changing (even slightly) the
main ingredients. This helps understanding the choices made below in discrete
settings.

The most popular and widespread form of the uncertainty principle uses the
variance as spreading measure of a function and its Fourier transform. This
leads to the inequality stated in Theorem~\ref{th:robertson}, where $A=X$ is the
multiplication operator $Xf(t)=tf(t)$ and $B=P=-i\partial_t/2\pi$ is the derivative
operator.  This first instance of uncertainty inequalities is associated to the
so-called canonical phase space, i.e. the time-frequency, or position-momentum
space. Let us first introduce some notations. Given $f\in\Ltwo$, denote by $\hat
f$ its Fourier transform, defined by
$$
\hat f(\nu) = \int\bds f(t) e^{-2i\pi\nu t}\,dt\ .
$$
With this definition, the Fourier transformation is an unitary operator
$\Ltwo\to\Ltwo$. The classical uncertainty inequalities state that for any
$f\in\Ltwo$, $f$ and $\hat f$ cannot be simultaneously sharply localized.

\paragraph{Heisenberg's inequality.} let $\cH=\Ltwo$ and consider
the self-adjoint operators $X$ and $P$, defined by $Xf(t)=t f(t)$ and
$Pf(t) = -i f'(t)/2\pi$. $X$ and $P$ satisfy the commutation relations
$[X,P]=i\,\Id$, where $\Id$ is the identity operator.
For $f\in\Ltwo$, denote by $e_f$ and $v_f$ its expectation and variance
(see {Theorem}~\ref{th:robertson}):
\begin{equation}\label{def:var}
e_f\!=\!e_f(X)\! =\! \frac1{\|f\|^2}\!\int\bds\!\!\!\! t |f(t)|^2\, dt\ ,\quad
v_f\!=\!v_f(X) \!=\! \frac1{\|f\|^2}\!\int\bds\!\!\!\! (t-e_f)^2 |f(t)|^2\, dt\ .
\end{equation}
Then the Robertson-Schr\"odinger inequality takes the form
\begin{corollary}
For all $f\in\Ltwo$,
\begin{equation}
v_f \cdot v_{\hat f}\ge \frac1{16\pi^2}\ ,
\end{equation}
with equality if and only if $f: t\to f(t) = a e^{-b(t-\mu)^2/2}$ is a Gaussian
function, up to time shifts, modulations, rescalings and chirping
($a,b,\mu\in\CC$, with $\Re(b)>0$).
\end{corollary}

\subsection{Variance time-frequency uncertainty principles on different  spaces} \label{diff:space}
Usual variance inequalities are defined for functions on the real line, or on
Euclidian spaces. It is important to stress that these inequalities do not
generalize easily to other settings, such as periodic functions, or more
general functions on bounded domains. First, the definition of mean and variance
themselves can be difficult issues\footnote{The definition and properties of the
  variance (and other moments) on compact manifolds is by itself a well defined
  field of research named {\em directional statistics}.}.
For example, the definition of the mean of a
function on the circle $S^1$ is problematic. Sticking to the above
notations, the operator $X$ is not well defined on $L^2(S^1)$ because of the
periodicity, the meaning of $e_f(X)$ is not clear,
and does definitely not represent the mean value of $f$. Adapted definitions of
mean and variance are required. For example, the case $\cH=L^2(M)$, where $M$ is
a Riemannian manifold, has been studied by various authors
(see~\cite{Erb11uncertainty} and references therein).  

For example, in the case
of the circle one definition of the mean value is given by $e_f=\arg\langle
f,Ef\rangle$  where $E\psi(t)=\exp(i2\pi t)\psi(t)$ (the so-called von Mises's
mean, see~\cite{Bre}). From this, an {\em angle-momentum} uncertainty
inequality has been obtained in~\cite{Judge63},~\cite{Bre}. Yet, additional difficulties appear: first the bound of the uncertainty principle is modified (compared to the
$\Ltwo$ case) and depends non trivially on the function $f$ involved. This implies that functions whose uncertainty product attains the bounds are not necessarily minimizers and the strict positivity of the lower bound may not be garanteed. 
The answer of the authors is to suggest to modify the definition of the variance (in addition to the modification of the mean).

Similar problems are encountered in various different situations, such as the
affine uncertainty which we account for below. All this shows that alternatives
to variance-based spreading measures are necessary. We will address these in
section~\ref{sec:cont.other} below.

\subsection{Different representations}
\label{sec:diff.rep}
In the Robertson-Schr\"odinger formulation, the two representation spaces under
consideration (which form the phase space) are $L^2$ spaces of the spectrum of
two self-adjoint operators $A$ and $B$. The spectral theorem establishes the
existence of two unitary maps $U_A$ and $U_B$ mapping $\cH$ to the two $L^2$
spaces; the images of elements of $\cH$ by these operators yield the two 
representations, for which uncertainty inequalities can be proven. It is worth
noticing that these representations can be (possibly formally) interpreted as
scalar products of elements of $\cH$ with (possibly generalized) eigenbases of
$A$ and $B$.

This allows one to go beyond the time-frequency representation and introduce
generalized phase spaces. We shall assume that the generalized phase space is
associated with self-adjoint operators $A_1,\dots A_k$, which are infinitesimal
generators of generalized translations, acting on some signal (Hilbert) space $\cH$.
Whenever two operators $A_j$, $A_l$ are such that there exists a unitary
transform $U$ which turn these two operators into the standard case (operator $X,
P$ defined above), one can obtain time-frequency type uncertainty inequalities.
In such cases, the lower bound is attained for specific choices of $f$, which
are the images of gaussian functions by the unitary transformation $U$. We will
refer to this construction as a canonization process. An example where
canonization is possible can be found in Remark~\ref{rem:klauder.vs.altes} 
below.

When this is not the case, the commutator $[A_j,A_l]$ is not a multiple of the
identity and the lower bound generally depends on $f$. This implies phenomena described in section~\ref{diff:space}. Even worse, if the spectrum of the 
operator $i[A_j,A_l]$ include zero then the lower bound is zero, revealing that the variance may not be a spreading measure in this case.

\subsubsection{Time-scale variance inequality}
The classical affine variance inequality is another particular instance of the
Robertson-Schr\"odinger inequality: let $A=X$ and $B=D = (XP+PX)/2$ denote
the infinitesimal generators of translations and dilations, acting on the Hardy
space $H^2(\RR) = \{f\in\Ltwo,\ \hat f(\nu)=0\ \forall\nu\le 0\}$,
which is the natural setting here.

Explicit calculation shows that $[X,D]=iX$, and it is worth introducing the {\bf
scale transform} $f\in H^2(\RR)\to\tf$, which is a unitary mapping
$\Htwo\leftrightarrow\Ltwo$ defined by
\begin{equation}
\tf(s) = \int_0^\infty \hat f(\nu) e^{2i\pi s}\frac{d\nu}{\sqrt{\nu}}
= \int\bds \hat f(e^u) e^{u/2} e^{2i\pi us}\, du\ .
\end{equation}
The corresponding Robertson-Schr\"odinger inequality states
\begin{corollary}
For all $f\in\Htwo$,
\begin{equation}
v_{\hat f}.v_{\tf}\ge \frac1{16\pi^2} e_{\hat f}^2\ ,
\end{equation}
with equality if and only if $\hat f$ is a {\em Klauder waveform}, defined by
\begin{equation}
\hat f(\nu) = K \exp\left\{a\ln(\nu) -b\nu + i (c\ln(\nu)+d)\right\}\ ,\qquad
\nu\in\RR^+
\end{equation}
for some constants $K\in\CC$, $a>-1/2$, $b\in\RR^+$ and $c,d\in\RR$.
\end{corollary}
It is worth noticing that the right hand side explicitely depends on $f$, so
that the Klauder waveform, which saturates this inequality, is not necessarily a
minimizer of the product of variances, as analyzed in~\cite{Maass10do}.

\subsubsection{Modified time-scale inequality}
The above remark prompted several authors (see~\cite{Flandrin01inequalities}
for a review) to seek different forms of averaging, adapted to the affine
geometry. This led to the introduction of adapted means and variances: for
$f\in\Htwo$, set
\begin{equation}
\tilde e_{\hat f} = \exp\left\{
\frac1{\|\hat f\|^2} \,
\int_0^\infty |\hat f(\nu)|^2\ln(\nu)\,d\nu\right\}\ ,\qquad
\tilde v_{\hat f} = \frac1{\|\hat f\|^2}\,
\int_0^\infty \left[\ln (\nu/\tilde e_{\hat f})\right]^2 |\hat f(\nu)|^2\,d\nu\ .
\end{equation}
In this new setting, one obtains a more familiar inequality
\begin{proposition}
For all $f\in\Htwo$,
\begin{equation}
\tilde v_{\hat f} . v_{\tf} \ge \frac1{16\pi^2}\ ,
\end{equation}
with equality if and only if $f$ takes the form of an {\em Altes waveform}, defined by
\begin{equation}
\hat f(\nu) = K \exp\left\{
-\frac1{2}\ln(\nu) - a \ln^2(\nu/b) + i (c\ln(\nu)+d)\right\}\ ,\qquad
\nu\in\RR^+\ ,
\end{equation}
which is now a variance minimizer.
\end{proposition}
\begin{figure}
\centerline{
\includegraphics[width=4cm]{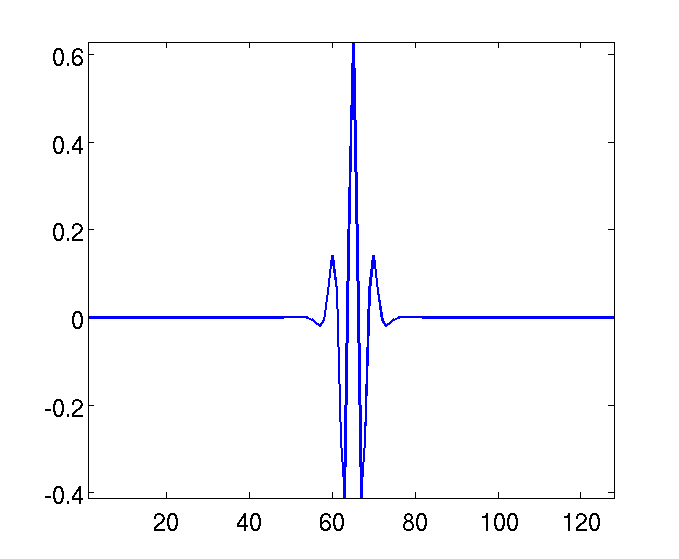}
\quad
\includegraphics[width=4cm]{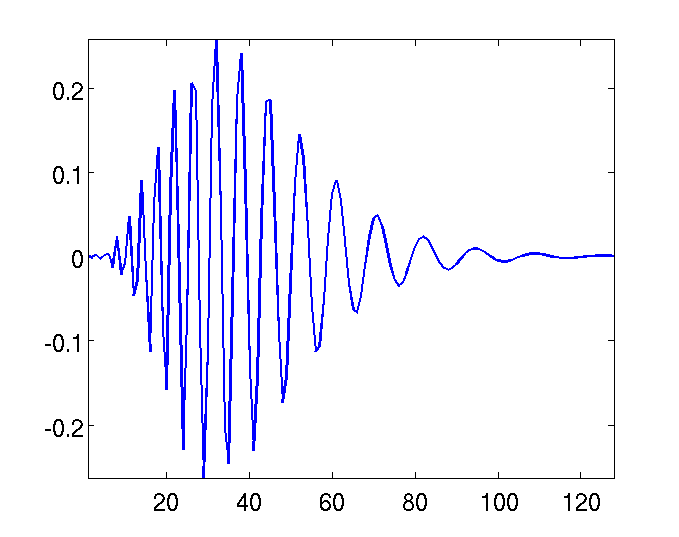}
}
\caption{Examples of Klauder waveform (left) and Altes waveform (right).}
\label{fi:waveforms}
\end{figure}

\begin{remark}[Canonization]
\label{rem:klauder.vs.altes}
The connection between Klauder's construction and Altes' can also be interpreted
in terms of canonization. Let $U: \Htwo\to\Ltwo$ denote the unitary linear
operator defined by $\widehat{Uf}(\nu) = e^{\nu/2}\hat f(e^\nu)$, for $\nu\in\RR_+$. The
adjoint operator reads $U^*f(s) = f(\ln(s))/\sqrt{s}$ (for $s\in\RR_+$),
and it is readily verified that $U$ is unitary.
Consider now the linear operators $\tX$ and
$\tP$ on $\Htwo$ defined by $\tX = U^*XU$ and $\tP = U^*PU$.
Simple calculations show that $\tX = D/2\pi$ and $\tP=2\pi\ln(P/2\pi)$, these
two operators being well defined on $\Htwo$. Hence $\tX$ and $\tP$ satisfy the
canonical commutation relations on $\Htwo$:
$$
[D,\ln(P)]=[D,\ln(P/2\pi)]=[\tX,\tP] = U^*[X,P]U = i\,\Id_\Htwo\ .
$$
Now, given any self adjoint operator $A$ on $\Htwo$, and for any $f\in\Htwo$,
set $g=Uf$, and one has $e_f(A)=e_g(UAU^*)$ and $v_f(A)=v_g(UAU^*)$. Therefore,
$$
v_f(D).v_f(\ln(P)) = v_g(X). v_g(P)\ge\frac1{16\pi^2}\ ,
$$
with equality if and only if $g$ is a Gaussian function, i.e. $f$ is an Altes
wavelet.
\end{remark}

\subsection{Different dispersion measures}
\label{sec:cont.other}
As stressed above, variance is not always well defined, and even when it is so,
variance inequalities may not yield meaningful informations. Alternatives have
been proposed in the literature, and we review some of them here.
Some of then show better stability to generalizations, and will be more easily
transposed to the discrete case.

\subsubsection{Hirschman-Beckner entropic inequality}
Following a conjecture by Everett~\cite{Everett57ManyWorlds} and
Hirschman~\cite{Hirschman57note}, Beckner~\cite{Beckner75inequalitieslong}
proved an inequality involving entropies. Assume $\|f\|=1$, and
define Shannon's differential entropy by
\begin{equation}
H(f) = \int |f(t)|^2\ln(|f(t)|^2)\, dt\ .
\end{equation}
Then the Hirschman-Beckner uncertainty principle states
\begin{theorem}
For all $f\in\Ltwo$,
\begin{equation}
H(f) + H(\hat f) \ge 1-\ln(2)\ ,
\end{equation}
with equality if and only if $f$ is a Gaussian function (up to the usual
modifications).
\end{theorem}
The proof originates from the {\em Babenko-Beckner
  inequality} (also called sharp Hausdorff-Young
inequality)~\cite{Beckner75inequalitieslong}: for $f\in L^p(\RR)$, let
$1/p+1/p'=1$; then $\|\hat f\|_{p'} \le A_p \|f\|_p$,
where $A_p=\sqrt{p^{1/p}/p'^{1/p'}}$. Taking logarithms after suitable
normalization yields an inequality involving Rényi entropies (see below for a
definition), that reduces to the Hirschman-Beckner inequality for $p=p'=2$.
As remarked in~\cite{Flandrin01inequalities}, for the time-scale uncertainty,
the canonization trick applies in this case as well, and yields a corresponding
entropic uncertainty inequality for time and scale variables.

\subsubsection{Concentration on subsets, the Donoho-Stark inequalities}
In~\cite{Donoho89uncertainty},
Donoho and Stark prove a series of uncertainty inequalities, in both continuous
and discrete settings, using different concentration measures. One of these is
the following: for $f\in\Ltwo$, and $\epsilon>0$, $f$ is said to be
$\epsilon$-concentrated in the measurable set $U$ if there exists $g$ supported
in $U$ such that $\|f-g\|\le\epsilon$. Donoho and Stark prove
\begin{theorem}
Assume that $f$ is $\epsilon_T$ concentrated in $T$ and $\hat f$ is
$\epsilon_F$ concentrated in $F$; then
\begin{equation}
|T| \cdot |F| \ge (1-(\epsilon_T+\epsilon_F))^2\ .
\end{equation}
\end{theorem}
\begin{remark}[Gerchberg-Papoulis algorithm]
This uncertainty inequality is used to prove the convergence of the
Gerchberg-Papoulis algorithm for missing samples restoration for
band-limited signals, as follows.
Let $F,T$ be bounded measurable subsets of the real line. 
Given $x\in\Ltwo$ such that ${\mathrm Supp}(\hat x)\subset F$, assume
observations of the form
$$
y(t) = \left\{\begin{array}{ll} x(t) + n(t)&\hbox{ if }t\not\in T\\
n(t)&\hbox{ otherwise}
\end{array}
\right.
$$
where $n$ is some noise, simply assumed to be bounded.

Denote by $P_T$ the orthogonal projection onto $L^2$ signals supported by $T$ in
the time domain, and by $P_F$ the corresponding projection in the frequency
domain.
If $|F|\cdot |T|<1$, then $\|P_TP_F\|<1$ and $x$ is stably recovered by solving
$$
\tilde x = (1-P_TP_F)\inv y\ ,
$$
where stability means $\|x-\tilde x\|\le C \|n\|$.
\end{remark}
The same paper by Donoho and Stark provides several other versions of the
uncertainty principle, in view of different applications.

\medskip
In a similar spirit, Benedicks theorem states that every pair of sets of
finite measure $(T,F)$ is {\em strongly annihilating}, i.e. there exists a
constant $C(T,F)$ such that for all $f\in\Ltwo$, 
\beq{fo:Jaming}
\|f\|_{L^2(\RR\backslash T)}^2 + \|\hat f\|_{L^2(\RR\backslash F)}^2\ge
\|f\|^2/C(T,F)\ .
\ee
We refer to~\cite{JamingNazarov} for more details, together with generalizations
to higher dimensions as well as explicit estimates for the constants $C(T,F)$.

\subsection{Non-separable dispersion measures}
\label{sec:cont.nponsep}
Traditional uncertainty principles bound joint concentration in two different
representation spaces. In some situations, it is possible to define a joint
representation space (phase space) and derive corresponding uncertainty
principles. This is in particular the case for time-frequency uncertainty. The
quantities of interest are then functions defined directly on the time-frequency
plane, such as the short time Fourier transform and the ambiguity function.
Given $f,g\in\Ltwo$, the STFT (Short time Fourier transform)
of $f$ with window $g$ and the ambiguity function of $f$ are respectively the
functions $\cV_gf, \cA_f\in L^2(\RR^2)$ defined by
\begin{equation}
\cV_gf(b,\nu) = \int\bds f(t) \overline{g}(t-b) e^{-2i\pi\nu t}\,dt\ ,
\quad \cA_f = \cV_f f\ .
\end{equation}
Concentration properties of such functions have been shown to be relevant in
various contexts, including radar theory (see~\cite{Lieb90integral}) or time-frequency
operator approximation theory~\cite{Doerfler11approximation}. We highlight a few
relevant criteria and results.

\subsubsection{$L^p$-norm of the ambiguity function: Lieb's inequality}
\label{se:cont.lieb}
E. Lieb (see~\cite{Brascamp76best} for example)
gives bounds on the concentration of the Ambiguity function (resp. STFT).
Contrary to Heisenberg type uncertainty inequalities, which privilege a
coordinate system in the phase space (i.e. choose a time and a frequency axis),
bounds on the ambiguity function don't. Here, concentration is measured by $L^p$
norms, and the bounds are as follows
\begin{theorem}
For all $f,g\in\Ltwo$,
\begin{equation}
\left\{
\begin{array}{ll}
\|\cA_f\|_p \ge B_p \|f\|_2^2&\hbox{ for }p<2\\
\|\cA_f\|_p \le B_p \|f\|_2^2&\hbox{ for }p>2\\
\|\cA_f\|_2 = \|f\|_2^2&
\end{array}
\right.\ ,
\qquad
\left\{
\begin{array}{ll}
\|\cV_gf\|_p \ge B_p \|g\|_2\|f\|_2&\hbox{ for }p<2\\
\|\cV_gf\|_p \le B_p \|g\|_2\|f\|_2&\hbox{ for }p>2\\
\|\cV_gf\|_2 = \|g\|_2\|f\|_2&
\end{array}
\right.
\end{equation}
where $B_p=(2/p)^{1/p}$ is related to the Beckner-Babenko constants.
\end{theorem}
The norm $\|\cdot\|_p$ can be regarded as a diversity,
or spreading measure for $p<2$ and as a sparsity, or concentration measure for
$p>2$ (see section~\ref{measures}). Again, the optimum is attained for Gaussian
functions. It is worth noticing that as opposed to the measures on subsets,
these concentration estimates are strongly influenced by the tail of the Gabor
transform or the ambiguity function. It is not clear at all that the latter is
actually relevant in practical applications. 

\subsubsection{Time-frequency concentration on compact sets}
As a consequence of Lieb's inequalities, one can show
(see~\cite{Grochenig01foundations} for a detailed account) the following 
concentration properties for ambiguity functions and STFTs: let
$\Omega\subset\RR^2$, measurable, and $\epsilon>0$ be such that
\begin{equation}
\int_\Omega |\cV_gf(b,\nu)|^2\,dbd\nu\ge (1-\epsilon) \|g\|^2\|f\|^2\ ,
\end{equation}
then
$\forall p>2$, $|\Omega|\ge (1-\epsilon)^{p/(p-2)} (p/2)^{2/(p-2)}$.
In particular, for $p=4$, this yields
\begin{equation}
|\Omega|\ge 2 (1-\epsilon)^2\ .
\end{equation}
\begin{remark}
It would actually be worth investigating possible corollaries of such estimates,
in the sense of Gerchberg-Papoulis. For instance, assume a measurable region $\Omega$
of a STFT has been discarded, under which assumptions can one expect to be able
to reconstruct stably the region ? Also, when $T$
is large, one can probably not expect much stability for the reconstruction,
however what would be reasonable regularizations for solving such a
time-frequency inpainting problem ? 
\end{remark}

\subsubsection{Peakyness of ambiguity function}
Concentration properties of the ambiguity function actually play a central role
in radar detection theory (see e.g.~\cite{Woodward80probability}). However, the
key desired property of ambiguity functions, namely {\em peakyness}, otherwise
stated the existence of a sharp peak at the origin, is hardly accounted for by
$L^p$ norms, entropies or concentration on compact sets as discussed above.

\begin{figure}
\centerline{
\includegraphics[width=5cm]{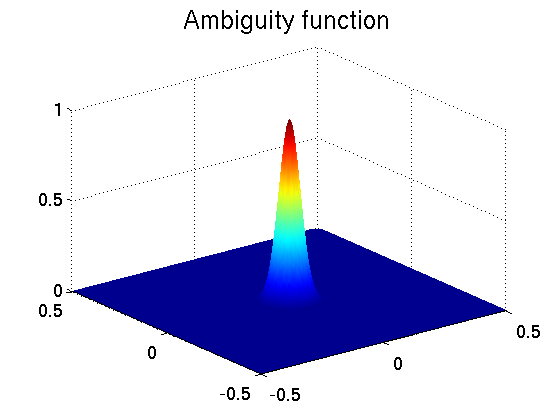}
\quad
\includegraphics[width=5cm]{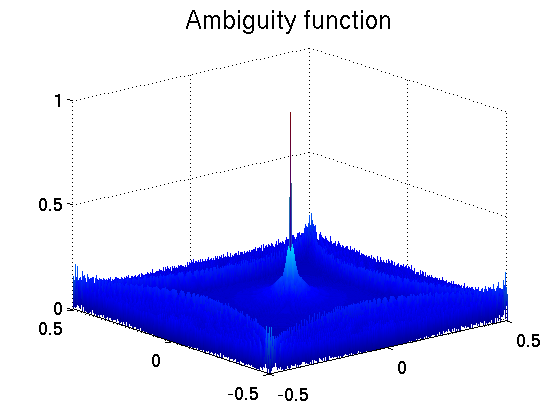}
}
\caption{3D plots of the ambiguity functions of a standard Gaussian (left) and a
  Hermite function of hight order (right).}
\label{fi:peakyness}
\end{figure}

Ambiguity function peakyness optimization can be formulated in a discrete
setting as follows. Suppose one is given a sampling lattice $\Lambda =
b_0\ZZ\times\nu_0\ZZ$ in the time-frequency domain, peakyness of $\cA_g$
can be optimized by maximizing (with respect to $g$) the quantity
\beq{fo:mu}
\mu(g) = \max_{(m,n)\ne(0,0)}\cA_g(mb_0,n\nu_0)\ .
\ee
Two examples of waveforms with different concentration properties are given in
Figure~\ref{fi:peakyness}. The gaussian function (left) has well known
concentration properties, while the ambiguity function of a high order hermite
function (right) is much more {\em peaky}, even though the function itself is
poorly time localized and poorly frequency localized.

The quantity in~\eqref{fo:mu} is actually closely connected (as remarked
in~\cite{Song10role}) to the so-called {\em coherence}, or {\em self-coherence}
of the Gabor family $\cD=\{g_{m,n},\,m,n\in\ZZ\}$ generated 
by time-frequency shifts $g_{mn}(t) = e^{2i\pi n\nu_0t} g(t-mb_0)$
of $g$ on the lattice $\Lambda$ (see~\cite{Grochenig01foundations} for a
detailed account), as
$$
\mu = \max_{(m',n')\ne (m,n)} |\langle g_{mn},g_{m'n'}\rangle|\ .
$$
Hence, optimizing the peakyness of the ambiguity function is closely connected
to minimizing the coherence of the corresponding Gabor family, a property which
has been often advocated in the sparse coding literature.
\begin{remark}
As mentioned earlier, sparsity requirements lead to minimize
the joint coherence in the case of separable uncertainty principles, and the
self-coherence in the case of non-separable uncertainty principles.
\end{remark}


\section{Discrete inequalities}
\label{se:discrete}

\subsection{Introduction}
\label{sec:sparsity}
The uncertainty principle in the discrete setting has gained increasing interest
during the last years due to its connection with sparse analysis and compressive
sensing. 
Sparsity has been shown to be an instrumental concept in various applications,
such as signal compression (obviously), signal denoising, blind signal separation,...
We first review here the main sparsity/diversity measures that have been used in
the signal processing literature, show that they are closely connected and
present several versions of the uncertainty principle. Then we present a few
examples of their adaptation to phase space concentration problems.

In the discrete finite-dimensional setting, we shall use the Hilbert space
$\cH=\CC^L$ as a model signal space. In terms of signal representations, we
consider finite frames $\cU=\{u_\lambda\in\cH,\,\lambda\in\Lambda\}$ (see
Section~\ref{signalrep} for motivations and  definitions) in $\cH$, and
denote by $U:x\in\cH\to\{\langle x,u_\lambda\rangle,\,\lambda\in\Lambda\}$ the
corresponding {\em analysis operator}, and by its adjoint $U^*$ the
{\em synthesis operator}.

The time-frequency frames offer a convenient and well established framework for
developing ideas and concepts, and most of the approaches described below have
been developed using Gabor frames. For the sake of completeness, we give here the
basic notations that will be used in the sequel. Given a reference vector
$\psi\in\cH$ (called the mother waveform, or the window), a corresponding Gabor
system associates with $\psi$ a family of time-frequency translates
$$
\psi_{mn}(t) = e^{2i\pi m\nu_0 t}\psi(t-nb_0)\ ,\quad
m\in\ZZ_M,\ n\in\ZZ_N,\ t\in\ZZ_L\ ,
$$
where $b_0=a/L$ and $\nu_0=b/L$ (with $a,b$ integers that divide $L$) are
constants that define a time-frequency lattice $\Lambda$. 
The corresponding transform $\cV_\psi$ associates with any
$x\in\cH$ a function $(m,n)\in\Lambda\to
\cV_\psi x(m,n) = \langle x,\psi_{mn}\rangle$.
When $a=b=1$, the corresponding transform is called the Short-time Fourier transform (STFT).

The ambiguity function of the window $\psi$ is the function $\cA_\psi$ defined
as the STFT of the waveform $\psi$ using $\psi$ as mother
waveform, in other words $\cA_\psi = \cV_\psi\psi$.

\subsection{Sparsity measures}\label{measures}

As mentioned earlier, the variance as a measure of spreading is problematic
in the finite setting both with its definition and the inequalities it yields
(see nevertheless~\cite{Nam13uncertainty} for an analysis of the connection
between continuous and finite variance inequalities).
More adapted measures have been proposed in the literature, among
which the celebrated $\ell^1$-norm used in optimization problems, entropy used by
physicists and in information theory and support measures favored for sparsity
related problems. 
\subsubsection{$\ell^p$-norms and support measure}
Given a finite-dimensional vector $x\in\CC^L$, it 
is customary in signal processing applications to use $\ell^p$ (quasi-) norms of
$x$ to measure the sparsity ($p>2$) or diversity ($p<2$) of the vector $x$:
\be
\|x\|_p = \sqrt[p]{\sum_{\ell=1}^L |x_\ell|^p}\ .
\ee
These quantities (except for $p=0$) do not fully qualify as sparsity or
diversity measures since they depend on the $\ell^2$-norm of $x$. To circumvent this
problem, normalized $\ell^p$-norms are also considered:
\be
\gamma_p(x) = \frac{\|x\|_p}{\|x\|_2} = \|\widetilde{x}\|_p\ ,
\text{ with } \widetilde{x}=x/\|x\|_2\ .
\ee
The normalized quantity $|\widetilde{x}|^2$ may be seen as a probability
distribution function.

The special case $p=0$ gives the support measure (number of non-zero
coefficients) also denoted $\ell^0$. This is not a norm but is obviously a
sparsity measure.

\subsubsection{R\'enyi entropies}
Entropy is a notion of disorder or spreading for
physicists and a well-established notion for estimating the amount of
information in information theory. Given $\alpha\in\RR^+$ and a vector
$x\in\CC^L$, the corresponding Rényi entropy~\cite{RenyiEntropy}
$R_\alpha(x)$ is defined as
\be
R_\alpha(x) = \frac{2\alpha}{1-\alpha}\ln \left(\gamma_{2\alpha}(x)\right)\
,\quad\alpha\ne 1\ .\label{eqRenyi}
\ee
R\'enyi entropies provide diversity measures, i.e. sparsity is obtained by
minimizing the entropies.
The limit $\alpha\to 1$ is not singular, and yields the Shannon entropy
\be
S(x) = -\sum_{\ell=1}^L
\frac{|x_\ell|^2}{\|x\|_2^2}\ln\left(\frac{|x_\ell|^2}{\|x\|_2^2}\right)\ .
\ee
These notions have been proven useful for measuring energy concentration in
signal processing, especially in the time-frequency
framework~\cite{Baraniuk01measuring} and~\cite{Jaillet07time}.

\subsubsection{Relations between sparsity measures}
Equation~\eqref{eqRenyi}
shows that minimizing the $\ell^p$-norm with $p<2$ is equivalent to minimizing
the R\'enyi entropy for $\alpha=p/2$. Note also that for $p=2\alpha>2$,
$1/(1-\alpha)$ is negative and minimizing the $\alpha$-entropy leads to the same
results as for \emph{maximizing} the $\ell^p$-norm. The limit $\alpha\to 1$
gives the Shannon entropy. Note also that the limit $\alpha=0$ is not singular
and gives the logarithm of the support size. Hence, all these measures a related
through Eq.~\eqref{eqRenyi} and belong to the same family.

So far, the focus has been put on the R\'enyi entropies and their limit, the
Shannon entropy; Tsallis entropies
$T_\alpha(x) = -(\gamma_{2\alpha}^{2\alpha}(x) -1)/(\alpha-1)$, initially
introduced in statistical physics, may be seen as some first order
approximations of R\'enyi entropies and can also be used along the same lines.
Comparison between these measures could be an interesting issue.

%

\subsection{Sparsity related uncertainties in finite dimensional settings}
\label{se:sparse}
Discrete uncertainty inequalities have received significant attention in many
domains of mathematics, physics and engineering. We focus here on the aspects
that have been mostly used in signal processing. 

\subsubsection{Support uncertainty principles}
The core idea is that in finite dimensional settings, two orthonormal bases
provide two different representations of the same object, and that the same
object cannot be represented sparsely in two ``very different bases''.
In the original work by Donoho and Huo~\cite{Donoho01uncertainty},
the finite-dimensional Kronecker and Fourier bases were used, and Elad and
Bruckstein~\cite{Elad02Generalized} extended the result to arbitrary orthonormal
bases.
The $\ell^0$ quasi-norm is used to measure diversity.
\begin{theorem}
Let $\Phi=\{\varphi_n,\,n\in\ZZ_N\}$ and $\Psi=\{\psi_n,\,n\in\ZZ_N\}$
denote two orthonormal bases of $\CC^N$. For all $x\in\CC^N$, denote by
$\alpha\in\CC^N$ and $\beta\in\CC^N$ the coefficients of the expansion of $x$ on
$\Phi$ and $\Psi$ respectively.
Then if $x\ne 0$
\be
\|\alpha\|_0 \cdot \|\beta\|_0 \ge \frac{1}{\mu^2}\quad\text{and}
\quad\|\alpha\|_0 + \|\beta\|_0 \ge \frac{2}{\mu}\ ,
\ee
where $\mu = \mu(\Phi,\Psi)$ is the mutual coherence
of $\Phi$ and $\Psi$ (see equation~\eqref{fo:mutual.coherence}).
\end{theorem}

\begin{remark}\rm
The {\em Welch bound} states that the mutual coherence of the union of two
orthonormal bases of $\CC^N$ cannot be smaller than $1/\sqrt{N}$; the bound is sharp,
equality being attained in the case of the Kronecker and Fourier bases.
\end{remark}
\begin{remark}\rm
The result was extended later on by Donoho and Elad~\cite{Donoho03Optimally} to
arbitrary frames, using the notion of Kruskal's rank (or spark): the Kruskal
rank of a family of vectors $\cD=\{\varphi_0,\dots\varphi_{N-1}\}$ in a
finite-dimensional space is the smallest number $r_K$ such that there exists a
family of $r_K$ linearly dependent vectors. 
Assume that $x\in\CC^N$, $x\ne 0$ has two different representations in
$\cD$:
$$
\hbox{if}\quad x=\sum_{n=0}^{N-1}\alpha_n\varphi_n=
\sum_{n=0}^{N-1}\beta_n\varphi_n\ ,
\qquad
\hbox{then}
\quad
\|\alpha\|_0 + \|\beta\|_0 \ge r_K\ .
$$
Bounds describing the relationship between the Kruskal rank and coherence have
also been given in~\cite{Donoho03Optimally}.
\end{remark}
Let us also mention at this point the discrete versions of the concentration
inequality~\eqref{fo:Jaming}, obtained in~\cite{GhobberJaming}. Given two bases
in $\CC^N$, let $T,F$ be two subsets of the two index sets $\{0,1\cdots,N-1\}$
and assume $|T|\cdot |F|<1/\mu^2$. Then for all $x$, 
\be
\|\alpha\|_{\ell^2(\ZZ_N\backslash T)}+\|\beta\|_{\ell^2(\ZZ_N\backslash F)}
\ge \left(1+\frac{1}{1-\mu\sqrt{|T|\cdot |F|}}\right)^{-1} \|x\|_2\
\ .
\ee
\begin{remark}
As expected, all these inequalities imply a strictly positive lower bound and
the coherence $\mu$. 
\end{remark}
In a recent study~\cite{RicaudTorresani} the support inequalities have been
extended from basis representations to frame ones.  More precisely, for any
vector $x\in\cH$, bounds of the following form have been obtained
\begin{theorem}
Let $\cU =\{\cU^{(1)},\dots\cU^{(n)}\}$ denote a set of $K$ frames in a Hilbert
space $\cH$. Then for any $x\in\cH$
\be
\sum_{k=1}^n \|U^{(k)}x\|_0 \ge \frac{n}{\mu_\star}\ ,
\ee
where $\mu_\star$ is a generalized coherence, defined as follows:
\beq{fo:mu.star}
\mu_\star = \inf_{\tcU}\inf_{1\le r\le 2}
\sqrt[n]{\mu_r(\tcU^{(1)},\cU^{(2)})\dots\mu_r(\tcU^{(n-1)},\cU^{(n)})
\mu_r(\tcU^{(n)},\cU^{(1)})}\ ,
\ee
where the infimum over $\tcU$ is taken over the family of all possible dual
frames  $\tcU =\{\tcU^{(1)},\dots\tcU^{(n)}\}$ of the elements of $\cU$, and the
r-coherences $\mu_r$ are defined as
\be
\mu_r(\cU,\cV) =\sup_{v\in\cV} \left(\sum_{u\in\cU} |\langle
u,v\rangle|^{r'}\right)^{r/r'}\ ,\quad \frac1{r} + \frac1{r'}=1\ .
\ee
\end{theorem}
Therefore, the control parameter here is the generalized coherence $\mu_\star$.
If the canonical dual frame is chosen, $\mu_r$ is often smaller than $\mu$ which
shows an improvement. This suggests new definitions for the coherence which may
improve further the inequality bound. 

\subsubsection{Entropic uncertainty}

In section~\ref{measures} we introduced the entropy as a measure of
concentration and we also stated earlier an entropic version of the uncertainty
principle for the continuous case (Hirschman-Beckner). It turns out that the
latter can be extended to more general situations than simply time-frequency
uncertainty. For example, in a discrete setting, given two orthonormal bases it
was proven by Maassen and Uffink~\cite{Maassen88generalized} and Dembo, Cover
and Thomas~\cite{Dembo91information} independently that for any $x$, the
coefficient sequences $\alpha$, $\beta$ of the two corresponding representations
of $x$ satisfy
\begin{equation}
S(\alpha)+S(\beta)\ge -2\ln\mu\ ,
\end{equation}
with $\mu$ the mutual coherence of the two bases.
In the particular case of Fourier-Kronecker bases, $\mu=1/\sqrt{N}$,
which leads to the similar result given in Prop.~\ref{discreteLieb} for
ambiguity function; the picket fences are the minimizers (see next section).  

\medskip
These results were generalized recently in~\cite{RicaudTorresani}, where
entropic inequality for frame analysis coefficients were obtained.
\begin{theorem}
\label{th:gen.entropic.ineq}
Let $\cH$ be a separable Hilbert space, let $\cU$ and $\cV$ be two frames of
$\cH$, with bounds $A_\cU,B_\cU$ and $A_\cV,B_\cV$.
Let $\tilde\cU$ and $\tilde\cV$ denote corresponding dual frames, and set
\beq{fo:notations}
\rho(\cU,\cV)=\sqrt{\frac{B_\cV}{A_\cU}}\ ,\quad
\sigma(\cU,\cV) = \sqrt{\frac{B_\cU B_\cV}{A_\cU A_\cV}}\ge 1\ ,\quad
\nu_r(\cU,\tcU,\cV) = \frac{\mu_r(\tcU,\cV)}{\rho(\cU,\cV)^r}\ .
\ee
Let $r\in [1,2)$.
For all $\alpha\in [r/2,1]$, let $\beta=\alpha(r-2)/(r-2\alpha)\in [1,\infty]$.
For $x\in\cH$, denote by $a$ and $b$ the sequences of analysis coefficient of
$x$ with respect to $\cU$ and $\cV$. Then
\begin{enumerate}
\item
The R\'enyi entropies satisfy the following bound:
\be
(2-r) R_\alpha(a) + rR_\beta(b) \ge -2\ln(\nu_r(\cU,\tilde\cU,\cV))
- \frac{2r\beta}{\beta-1}\ln(\sigma(\cU,\cV))
\ee
\item
If $\cU$ and $\cV$ are tight frames, the bound becomes
\be
(2-r)R_\alpha(a) +r R_\beta(b)\ge -2\ln(\nu_r(\cU,\tcU,\cV))\ .
\ee
\item
In this case, the following inequalities between Shannon entropies hold true:
\be
S(a)+S(b) \ge -2\ln\left(\mu_\star(\cU,\tcU,\cV,\tilde\cV)\right)\ ,
\ee
where $\mu_\star$ is defined in~\eqref{fo:mu.star}.
\end{enumerate}
\end{theorem}
The proof is both a refinement and a frame generalization of the proof
in~\cite{Maassen88generalized,Dembo91information}. A main result
of~\cite{RicaudTorresani} is the fact that these (significatly more complex)
bounds indeed provide stronger estimates than the Maassen-Uffink inequalities,
even in the case of orthonormal bases. They are however presumably sub-optimal
for non tight frames, as they yield in some specific limit support inequalities
that turn out to be weaker than the ones presented above.

\subsubsection{Phase space uncertainty and localization}
\label{se:discr.lieb}
Again, as in the continuous case, uncertainty inequalities defined directly in
phase space can be proven. For example, in the joint time-frequency case,
finite-dimensional analogues of Lieb's inequalities have been proven
in~\cite{Optimwin}. 

\begin{figure}
\centerline{
\includegraphics[width=6cm]{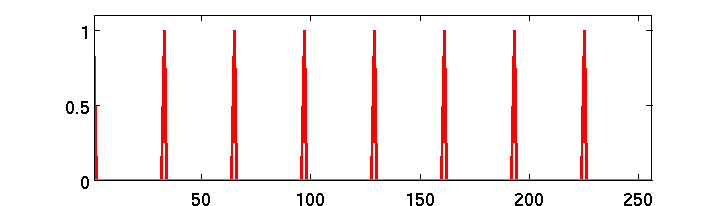}
\includegraphics[width=6cm]{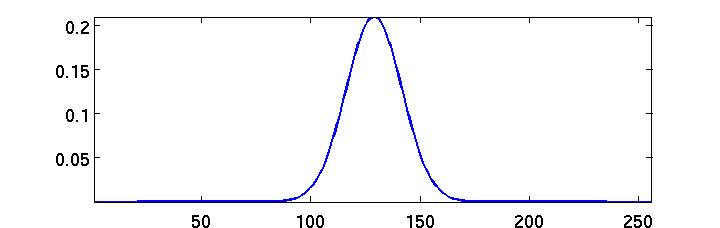}
\qquad
}
\caption{Picket fence (left) vs periodized gaussian (right)}
\label{fi:picket.gaussian}
\end{figure}

\begin{proposition}\label{discreteLieb}
Let $\psi\in\CC^N$ be such that $\|\psi \|_2=1$. Then, assuming $p<2$,
\be
\label{fo:discrete.Lieb}
\|\cA_\psi \|_p\ge N^{\frac1{p}-\frac1{2}}\ ,\qquad\hbox{and}\quad
S(\cA_\psi )\ge\log(N)\ .
\ee
The inequality is an equality for the family of ``picket fence'' signals,
translated and modulated copies of the following periodic series of Kronecker deltas: 
$$
\omega(t)=\frac{1}{\sqrt{b}}\sum_{n=1}^{b} \delta(t-an),\quad ab=N\ .
$$
\end{proposition}
Hence, the result is now completely different from the result obtained in the
continuous case: The optimum is not the Gaussian function (which by the way is
not well defined in finite-dimensional situations) any more, and is now a
completely different object, as examplified in Fig.~\ref{fi:picket.gaussian},
where a picket fence and a periodized Gaussian window are displayed. This is
mainly due to the choice of underlying model signal spaces (generally $\Ltwo$),
which impose some decay at infinity.

\begin{remark}It is worth noticing that the above diversity measures (norms or
  entropy of the ambiguity function) are non-convex functionals of the window
  sequence. For example, if $N$ is a prime number, there are (up to
  normalization) $2N$ window vectors (picket fences) whose ambiguity function is
  optimally concentrated (in terms of entropy). When $N$ is not prime, the
  degeneracy is even higher. 
\end{remark}

\subsection{Two signal processing applications} 

The uncertainty principle and its consequences have long been considered as
constraints and barriers to access precise knowledge and measurement. The
innovative idea behind compressive sensing where the uncertainty principle is
turned into an advantage for retrieving information promise many exciting
developments. In this section we present two prototype applications which
involve the uncertainty principle. In the same spirit as compressive sensing,
the first one shows how the uncertainty principle can be used for the separation
of signals. The second application is a more classical one which provides
time-frequency windows with minimum uncertainty, under some additional
constraints.

\subsubsection{Sparsity-based signal separation problem}

The signal separation problem is an extremely ill-defined signal processing
problem, which is also important in many engineering problems. In a
nutshell, it consists in splitting a signal $x$ into a sum of {\em components}
$x_k$, or {\em parts}, of different nature:
$$
x=x_1+x_2+\dots +x_n\ .
$$
While this notion of {\em different nature}
often makes sense in applied domains, it is generally extremely difficult to
formalize mathematically. Sparsity (see~\cite{Kutyniok12data} for an
introduction in the data separation context) offers a convenient framework for
approaching such a notion, according to the following paradigm:

\noindent{\em Signals of different nature are sparsely represented in different waveform
systems.
}

Given a union of several frames (or frames of subspaces)
$\cU^{(1)},\cU^{(2)},\dots \cU^{(n)}$ in a reference Hilbert space $\cH$,
the separation problem can be given various formulations, among which the
so-called analysis and synthesis formulations.
\begin{itemize}
\item
In the {\em synthesis formulation}, each component $x_k$ will be synthesized using the
$k$-th frame in the form $\sum_j \alpha_j^{(k)} u_j^{(k)}$, and the synthesis
coefficients $\alpha$ will be sparsity constrained. The problem is then settled
as
$$
\min \sum_{k=1}^n \|\alpha^{(k)}\|_0\ ,\quad\hbox{under constraint}\quad
x=\sum_{k=1}^n \sum_j \alpha_j^{(k)} u_j^{(k)}\ .
$$
\item
In {\em analysis formulations} the splitting of $x$ is sought directly as the
solution of 
$$
\min_{x_1,\dots x_n\in\cH} \sum_{k=1}^n \|U^{(k)}x_k\|_0\ ,
\quad\hbox{under constraint}\quad
x=x_1+x_2+\dots x_n\ ,
$$
where $U^{(k)}$ denotes the analysis operator of frame $k$.
\end{itemize}
\bigskip

In the case of two frames, it may be proven that if one is given a splitting
$x=x_1+x_2$, obtained via any algorithm, if $\|U^{(1)}x_1\|_0 +
\|U^{(2)}x_2\|_0$ is small enough, then this
splitting is necessarily optimal. More precisely~\cite{RicaudTorresani}
\begin{corollary}
Let $\cU^{(1)}$ and $\cU^{(2)}$ denote two frames in $\cH$. For any $x\in\cH$, let
$x=x_1+x_2$ denote a splitting such that
$$
\|U^{(1)}x_1\|_0 + \|U^{(2)}x_2\|_0 <\frac1{\mu_\star}\ .
$$
Then this splitting minimizes $\|U^{(1)}x_1\|_0 + \|U^{(2)}x_2\|_0$.
\end{corollary}
Hence, the performances of the analysis-based signal separation problems rely
heavily on the value of this generalized coherence function.

The extension to splittings involving more than two parts is more cumbersome. It
can be attacked recursively, but this involves combinatorial problems which are
likely to be difficult to solve.

\subsubsection{Sparsity-based algorithms for window optimization in
  time-frequency analysis} 
\label{sec:optim}
Proposition~\ref{discreteLieb} shows that the finite dimensional waveforms that
optimize standard sparsity measures in the ambiguity domain are not localized,
neither in time nor in frequency. This was also confirmed by numerical
experiments reported in~\cite{Optimwin}, where numerical schemes for ambiguity
function optimization were proposed. This approach has so far been
developed mainly with time-frequency representations, but is generic
enough to be adapted to various situations.

More precisely, the problem addressed by these algorithms is the following: solve
\begin{equation}
\label{varamb2}
\psi_{\rm opt}=\underset{\psi : \|\psi \|=1}{\rm arg\ max} \sum_z F(|\cA_\psi (z)|,z)
\left|\cA_\psi (z)\right|^2\ ,
\end{equation}
for some {\em density function} $F:\R^+\times \Lambda\to\R^+$, chosen so as to
enforce some specific localization or sparsity properties.
A simple approach, based upon quadratic approximations of the target
functional, reduces the problem to iterative diagonalizations of Gabor
multipliers. 

\begin{figure}[ht]
\begin{center}
\includegraphics[width=.4\textwidth]{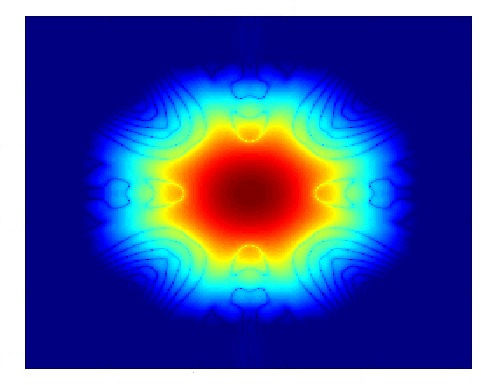}
\includegraphics[width=.4\textwidth]{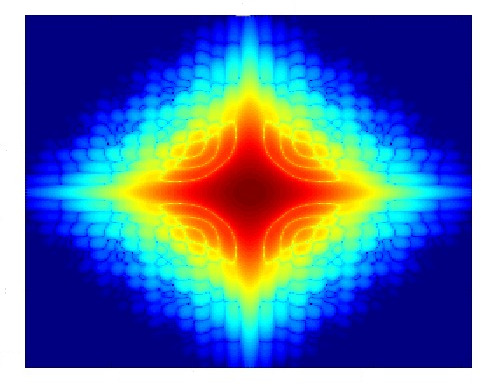}
\caption{Logarithm of modulus of optimal ambiguity functions with mask
  $F(|\cA_{\psi}(z)|,z)=F(0,z)$.
Left: Optimal function obtained for $F$ the indicator of a disk..
Right: Optimal function obtained for $F$ the indicator of a diamond.
}\label{fig:Amb}
\end{center}
\end{figure}

Two specific situations were considered and analyzed, namely:
\begin{itemize}
\item
the optimization of the ambiguity function sparsity through the maximization of
some $\ell^p$ norm (with $p>2$), which naturally leads to choose
$F(|\cA_\psi(z)|,z) = |\cA_\psi (z)|^{p-2}$.
The functional to optimize is non-convex, and the outcome of the
algorithm depends on the initialization. In agreement with the theory, numerical
experiments can converge to picket fence signals (Dirac combs) as limit windows. In
addition, for some choices of the initial input window, a Gaussian-like
function (the Gaussian is the sparsest window in the continuous case) may also
be obtained (local minimum).  
\item
the optimization of the concentration within specific regions, through choices
such as $F(|\cA_\psi (z)|,z)=F_0(z)$, for some non-negative function $F_0$
satisfying symmetry constraints, due to the particular
properties of the ambiguity function ($\cA(0,0)=1$, $\cA(z)=\cA(-z)$). The
algorithm were shown to converge to optimal windows matching the 
shape of $F$ in the ambiguity plane. That is to say this window is sharply
concentrated \emph{and} satisfy the shape constraint provided by $F$. However,
the convergence is not guaranteed for all $F$ and convergence issues should be
treated in more details in future works. The algorithm has been shown to
converge for simple shapes such as discs, ellipses or rectangles in the
ambiguity plane. Numerical illustrations can be found in Fig.~\ref{fig:Amb}
(disc shape and rectangular/diamond shape). Since the Ambiguity plane is
discrete, the masks are polygons rather that perfect circle and diamonds, and
this implies the amazing shape of the ambiguity function, with interferences.
For some more complex
shapes (such as stars for examples), the algorithm was found not to converge;
convergence problems are important issues, currently under study.
\end{itemize}

Such approaches are actually fairly generic, and there is hope that they can be
generalized so as to be able to generate waveforms that are optimal with respect
to large classes of criteria.

\section{Conclusions}
\label{se:Conclusions}
We have reviewed in this paper a number of instances of uncertainty
inequalities, in both continuous and discrete situations. 
Through these particular examples we have focused on specific properties and
connections between these different instances. 
Indeed, from its first statement in quantum mechanics to its newest developments
in signal processing, the uncertainty principle has encountered many parallel
evolutions and generalizations in different domains.
It was not a smooth and straightforward progress, as
different situations call for adapted spreading measures, yield different
inequalities, bounds and different minimizers (if any), and involve different
proof techniques.
A main point we have tried to make in this paper is that several classical
approaches, developed in the continuous setting, do not go through in more
general situations, such as discrete settings. For example, the very notions of
mean and variance do not necessarily make sense in general. In such situations
other, more generic, spreading measures such as the (R\'enyi) entropies and
$\ell^p$-norms can be used. We attempted in this paper to point out the close
connection between these quantities and suggest other candidates for further research.

Signal representations were first understood as the function itself and its
Fourier transform. It was then generalized to any projection on orthonormal
bases and now any set of frame coefficients. These latter representations play
an important role in signal processing and bring some new insight on the
uncertainty bounds. The introduction of the mutual coherence measuring how close
two representations can be, as well as the phase space coherence that measures
the redundancy of a corresponding waveform system, lead to new corresponding
bounds. A careful choice for this quantity is needed for obtaining the sharpest
bound possible. We showed how this notion of coherence can be extended and
generalized, using $\ell^p$-norms with $p\neq\infty$.

Concerning the uncertainty optimizers, i.e. waveforms that optimize an uncertainty
inequality, they are of very different nature in the discrete and continuous cases.
In a few words, in the continuous situations, some underlying choice of
functional space implies localization as a consequence of concentration (as
measured by the chosen spreading criterion). This is no longer the case in the
discrete world where localization and concentration have different meanings.

Therefore, the transition from continuous to discrete spaces is far more complex
than simply replacing integrals by sums and a more thorough analysis of the
connections between them is clearly needed.

%

\bibliographystyle{abbrv}
\bibliography{benjbib,TechRep1,UnLocX,addbib}   


\end{document}